\documentclass[global,twocolumn,final]{svjour}
\usepackage{graphics}
\journalname{Applied Physics B}

\begin{document}

\title{Power-efficient frequency switching of a locked laser}
\author{R.A. Cornelussen \and T.N. Huussen \and R.J.C. Spreeuw \and and H.B. van Linden van den Heuvell}
\institute{Van der Waals - Zeeman Instituut, Universiteit van Amsterdam, Valckenierstraat 65, 1018 XE  Amsterdam, The
Netherlands \email{ronaldc@science.uva.nl}}


\maketitle

\abstract{We demonstrate a new and efficient laser-locking technique that enables making large frequency jumps while
keeping the laser in lock. A diode laser is locked at a variable offset from a Doppler-free spectral feature of
rubidium vapor. This is done by frequency shifting the laser before sending the light to a spectroscopy cell with an
acousto-optic modulator (AOM). The frequency of the locked laser is switched quasi-\-instantaneously over much more
than the width of the spectral features, i.e. the usual locking range. This is done by simultaneously switching the AOM
frequency and applying feed-forward to the laser current. The advantage of our technique is that power loss and beam
walk caused by the AOM do not affect the main output beam, but only the small fraction of light used for the
spectroscopy. The transient excursions of the laser frequency are only a few MHz and last approximately 0.2 ms, limited
by the bandwidth of our locking electronics. We present equations that describe the transient behavior of the error
signal and the laser frequency quantitatively. They are in good agreement with the measurements. The technique should
be applicable to other types of lasers.}

\section{Introduction}
\label{sec1}

For the purpose of nowadays ubiquitous laser-cooling experiments \cite{cite1,cite2,cite3} lasers are routinely locked
to a Doppler-free absorption feature of an atomic transition. In a typical experimental time sequence one would first
accumulate atoms into a magneto-optical trap (MOT) followed by a phase of e.g. polarization gradient cooling. Both
phases require different detunings of the laser light. The required switching of the frequency has been solved in
several ways. However the existing solutions have some disadvantages, especially in terms of efficiency of laser power.

An existing and straightforward method is to lock the laser to a fixed detuning away from resonance and shift the laser
frequency towards resonance by a variable amount, using an acousto-optic modulator (AOM). Usually the AOM is used in
double-pass configuration to cancel beam walk associated with frequency shifting, resulting in a limited efficiency,
typically lower than 65\%. Moreover the beam walk compensation is imperfect.

Both the loss of laser power and the residual beam walk can be a problem when the light is used directly in an
experiment, or when high-power multi-mode amplifier lasers are used, such as a broad-area laser (BAL)
\cite{cite4,cite5} or a semiconductor tapered-amplifier laser (TA) \cite{cite6}. With more seeding power such
amplifiers perform better in terms of spectral purity and output power. Moreover they impose strict requirements on the
beam pointing stability of the injection beam. The latter problem could be solved by first amplifying the light before
frequency shifting it, providing the amplifier with sufficient power and a stably aligned injection beam. However power
loss and beam walk now occur in the amplified beam.

Another solution is to injection-lock \cite{cite7,cite8,cite9} a second single-mode diode laser with the frequency
shifted light, since this puts less stringent requirements on injection power and beam pointing stability. This
injection-locked diode laser can subsequently be used to seed a BAL or TA laser \cite{cite5}. A drawback is that this
solution requires a significant amount of extra equipment. Furthermore, it is not possible to implement this solution
in commercial BAL or TA systems without making major adjustments to the system, because the master laser is integrated
in the system.

\begin{figure}[t]
\centerline{\scalebox{.2}{\includegraphics{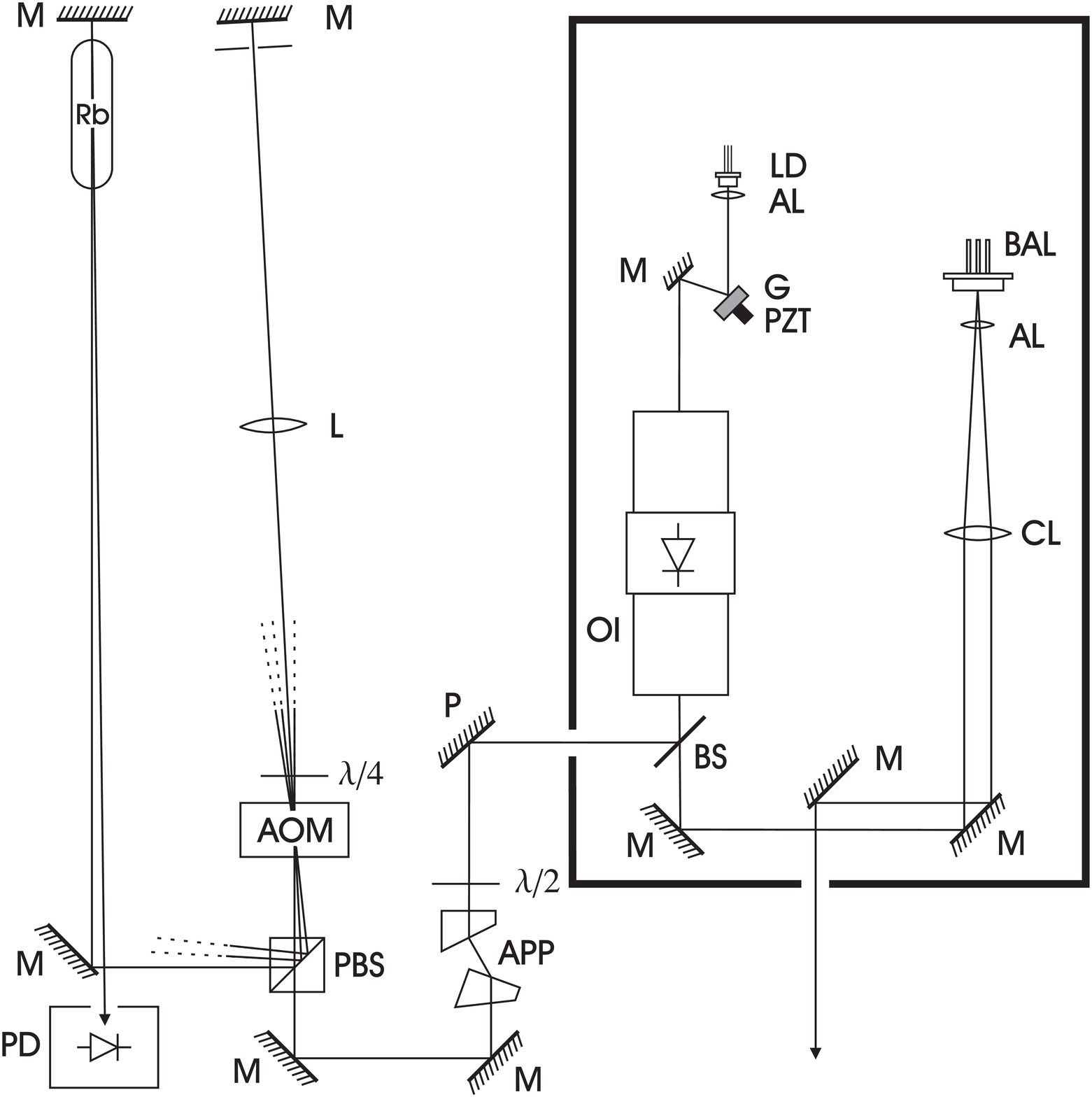}}} \caption{Schematic representation of the spectroscopy setup. The
spectroscopy beam is sent through a double pass AOM setup to a saturated absorption section. This allows locking of the
laser to an arbitrary frequency in the vicinity of an atomic transition. AL: aspheric lens, AOM: acousto-optic
modulator, APP: anamorphic prism pair, BAL: broad area laser, BS: beam splitter, CL: cylinder lens, G: grating, L:
lens, LD: laserdiode, M: mirror, OI: optical isolator, P: periscope, PBS: polarizing beam splitter, PD: photodiode,
PZT: piezo transducer, Rb: cell with rubidium vapor.} \label{fig1}
\end{figure}

In this paper we demonstrate our method which is both efficient in its use of laser power and rigorously eliminates the
beam walk due to the AOM frequency switching. We lock our laser using Doppler-free saturation spectroscopy in a vapor
cell of rubidium. The laser is frequency-shifted by an AOM before sending it through the spectroscopy cell. Thus,
instead of shifting a fixed-frequency laser by a variable amount, we lock the laser at a variable frequency. This is
only possible if the laser can follow the change in lock point associated with a change in the AOM frequency. This is a
problem if the frequency jump is larger than the locking range set by the width of the Doppler-free features in the
spectrum. We solved this by providing the laser with a feed-forward signal, causing the laser to jump to within the
locking range of the shifted lock point. We analyze the transient behavior of the laser frequency when making these
jumps.

\section{Experimental implementation}
\label{sec2}

In our experiment we work with $^{87}$Rb which has a natural linewidth of $\Gamma/2\pi$ = 6 MHz on the
$5\mathrm{S}_{1/2}\rightarrow 5\mathrm{P}_{3/2}$ resonance line (D2, 780 nm). The laser detunings needed for the MOT
and the molasses phase are -1.5$\Gamma$ and -10$\Gamma$ with respect to the $F=2\rightarrow F'=3$ component of the D2
line. In view of the frequency range of our AOM we lock the spectroscopy beam to the $F=2\rightarrow F'=(1,3)$
crossover. The detunings with respect to this transition are 203 MHz and 152 MHz, respectively. The desired frequency
jump of $\sim 50$ MHz is thus much larger than the locking range of about $\Gamma/2\pi$.

We use a commercial laser system (Toptica, PDL100) consisting of an extended cavity diode laser \cite{cite10,cite11} in
Littrow configuration \cite{cite12}, which injection locks a BAL. The grating of the extended cavity is mounted on a
piezo stack (PZT) in order to scan the frequency. The setup is shown in Fig. \ref{fig1}. Behind the 60 dB optical
isolator, 35~mW of power is left. The beam splitter reflects 10\% to the spectroscopy setup. The spectroscopy beam
first passes an anamorphic prism pair to circularize the elliptic beam shape. It then goes to a double pass AOM setup
and finally to a Doppler-free spectroscopy section.

\begin{figure}[t]
\centerline{\scalebox{.2}{\includegraphics{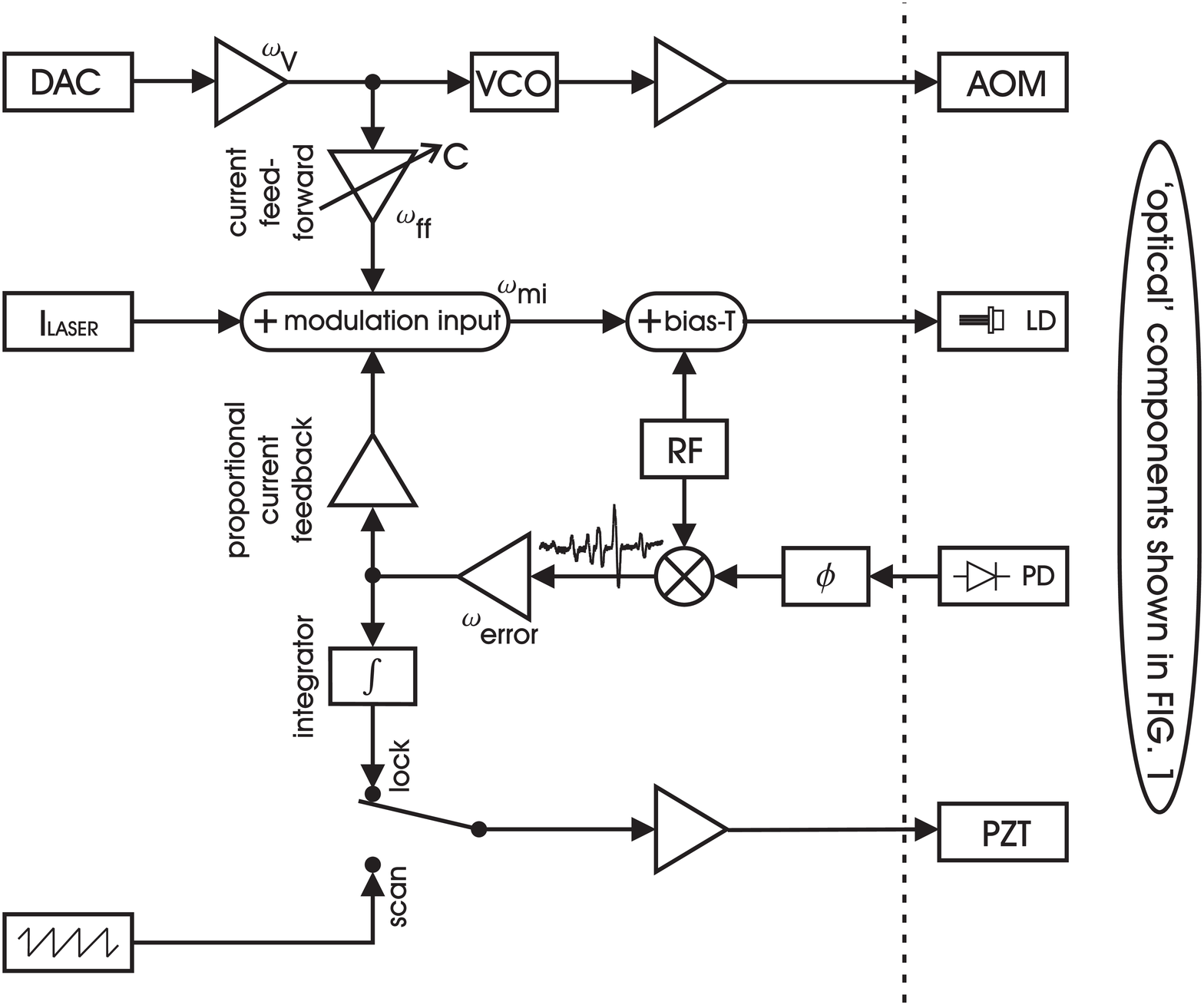}}} \caption{Schematic representation of the locking electronics. We
employ FM spectroscopy to obtain a dispersive signal. This signal is used for integrational feedback to the PZT for
proportional feedback to the laser current. The AOM frequency is generated by a VCO. The voltage driving the VCO is
amplified and used as feed-forward to the laser current in order to compensate the spectroscopy frequency for the
frequency change of the AOM. On the left side of the dashed line the locking electronics is shown, on the right side
the 'optical' components that can also be found in Fig. \ref{fig1} are visible.} \label{fig2}
\end{figure}

Fig. \ref{fig2} shows a schematic representation of the electronics to lock the laser frequency. We employ FM
spectroscopy \cite{cite13,cite14} to lock the laser. A small modulation with a radio-frequency (RF) of 33 MHz is added
to the laser current by means of a bias-T. The photodiode signal of the Doppler-free spectroscopy is phase shifted and
mixed with the RF frequency resulting in a dispersive error signal, which is amplified with a measured bandwidth
$\omega_{\mathrm{error}}/2\pi\approx 20$ kHz. This signal is integrated and sent to the PZT in order to lock the laser
to a spectral feature. Proportional current feedback is also applied to suppress fast fluctuations of the laser
frequency. The AOM frequency is generated by a voltage controlled oscillator (VCO). The voltage driving the VCO is
generated by a 12 bit digital to analog convertor (DAC), which is subsequently converted to the correct voltage range.
This last step has a measured bandwidth $\omega_{\mathrm{V}}/2\pi=2.6$ kHz.

\begin{figure}[t]
\centerline{\scalebox{.6}{\includegraphics{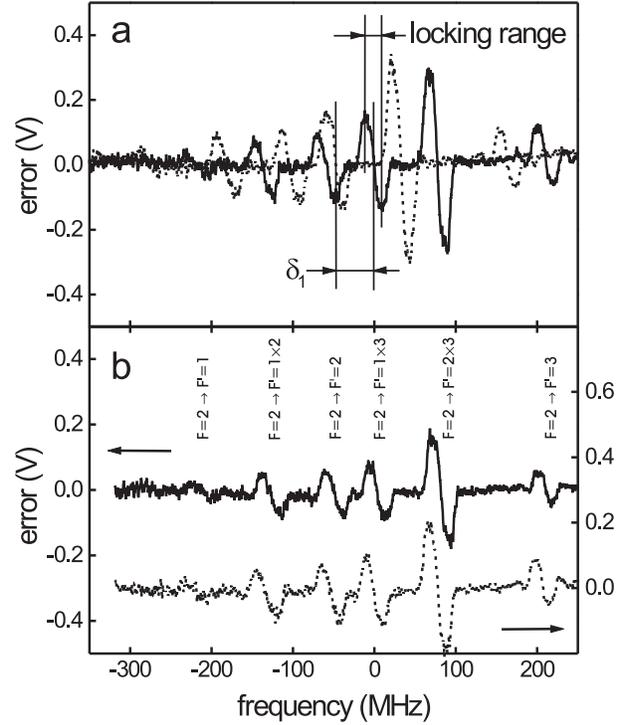}}} \caption{FM spectra, measured by scanning the PZT, for two
different AOM detunings (a) without and (b) with current feed-forward. It is clear that the frequency jump $\delta_{1}$
is larger than the locking-range, the width of the dispersive features. The laser will thus not stay locked to the $F=2
\rightarrow F'=(1,3)$ cross-over when the AOM frequency is changed without current feed-forward. Note that the curves
in (b) have been displaced vertically relative to each other.} \label{fig3}
\end{figure}

Fig. \ref{fig3}(a) shows FM spectra measured by scanning the PZT for two AOM detunings $\delta_{\mathrm{AOM}}$, which
are close to the frequencies used in a typical lasercooling experiment: $\delta_0/2\pi=-186.2$ MHz and
$(\delta_0+\delta_1)/2\pi=-137.4$ MHz. The jump in frequency is clearly larger than the half-width of the dispersive
features, so that the locking electronics will not be able to keep the laser locked to the same line when this
frequency jump is made. When this shift is compensated by applying a feed-forward jump to the laser current the
spectroscopy beam will not change frequency and the laser will stay locked. Experimentally this is done by attenuating
the voltage driving the VCO (measured bandwidth of the attenuator $\omega_{\mathrm{ff}}/2\pi=125$ kHz) and feeding this
as feed-forward to the modulation input of the current controller, which has a specified bandwidth
$\omega_{\mathrm{mi}}/2\pi=100$ kHz. Ideally the frequency change due to feed-forward $\delta_{\mathrm{ff}}$ and the
AOM detuning $\delta_{\mathrm{AOM}}$ should cancel. In reality the two frequencies are only approximately equal:
\begin{equation}
\delta_{\mathrm{ff}}=-C \delta_{\mathrm{AOM}}, \label{eq1}
\end{equation}
with $C\approx 1$. The parameter $C$ is coarsely adjusted to 1 by optimizing the overlap of the two spectra. The
accuracy is limited by the noise on the curves. Spectra with $C$ adjusted to 1 by this method are shown in Fig.
\ref{fig3}(b). When using the feed-forward on the laser current, we observe that the laser remains locked while
jumping. In the next section the transient behavior of the error signal will be discussed. A more accurate method to
optimize $C$ will be demonstrated in section \ref{sec4}.

\section{Analysis of transient behavior}
\label{sec3}

In this section an equation will be derived describing the transient behavior of the error signal. When the laser is
locked to a dispersive spectral feature and the frequency excursions are small with respect to the width of this
feature, the error signal $e(t)$ can be approximated by
\begin{equation}
e(t)=A(\omega_{\mathrm{S}}(t)-\omega_{\mathrm{R}}), \label{eq2}
\end{equation}
with $A$ the slope of the dispersive signal of the reference feature at frequency $\omega_{\mathrm{R}}$, which is equal
to the $F=2 \rightarrow F'=(1,3)$ cross-over frequency in our experiment. The frequency $\omega_{\mathrm{S}}(t)$ of the
light in the spectroscopy section is given by
\begin{equation}
\omega_{\mathrm{S}}(t)=\omega_{\mathrm{L}}(t)+\delta_{\mathrm{AOM}}(t), \label{eq3}
\end{equation}
where $\omega_{\mathrm{L}}(t)$ is the laser frequency and $\delta_{\mathrm{AOM}}(t)$ is the shift in the double-pass
AOM section. In our experiment this is a step function
\begin{equation}
\delta_{\mathrm{AOM}}(t)=\delta_0+\delta_1 u(t), \label{eq4}
\end{equation}
with $\delta_0/2\pi=-186.2$ MHz, $\delta_1/2\pi=48.8$ MHz and $u(t)$ the unit step function, so that the laser changes
frequency at $t=0$. Including all feedback and feed-forward terms the laser frequency $\omega_{\mathrm{L}}(t)$ is given
by
\begin{equation}
\omega_{\mathrm{L}}(t)=\omega_0+\varphi e(t)+\alpha\int_{-\infty}^{t}e(\tau)\mathrm{d}\tau+\delta_{\mathrm{ff}}(t).
\label{eq5}
\end{equation}
Here $\omega_0$ is the frequency of the laser when it is not locked or any other electronic feedback is applied, the
second term represents proportional current feedback. The third term is the integrational feedback to the PZT
controlled grating. The last term is the feed-forward to the laser current, which should instantaneously compensate the
detuning jump by the AOM, as defined in Eq. (\ref{eq1}).

When the laser is locked at $t=0$, before the frequency jump, several terms cancel:
\begin{equation}
\omega_{\mathrm{L}}(0_{-})=\omega_0+\varphi
e(0_{-})+\alpha\int_{-\infty}^0e(\tau)\mathrm{d}\tau-C\delta_0=\omega_{\mathrm{R}}-\delta_0. \label{eq6}
\end{equation}
Combining Eqs. (\ref{eq1})-(\ref{eq6}) yields:
\begin{equation}
e(t)=A\left(\varphi e(t)+\alpha\int_{0}^{t}e(\tau)\mathrm{d}\tau+(1-C)\delta_1 u(t)\right) \label{eq7}
\end{equation}
from which the error function $e(t)$ after the frequency jump can be solved. As discussed in the previous section,
several of the components have a limited bandwidth, which can be easily incorporated in the Laplace transform of Eq.
(\ref{eq7}), yielding
\begin{eqnarray}
E(s)=\tau_{\mathrm{error}}(s)A\times\left[\left(\tau_{\mathrm{mi}}(s)\varphi+\frac{\alpha}{s}\right)E(s)+\right.\nonumber\\
\left.\tau_{\mathrm{V}}(s)\left(1-\tau_{\mathrm{mi}}(s)\tau_{\mathrm{ff}}(s)C\right)\frac{\delta_1}{s}\right],
\label{eq8}
\end{eqnarray}
with $E(s)$ the Laplace transform of $e(t)$ and $\tau_{\mathrm{x}}(s)=1/(1+s/\omega_{\mathrm{x}})$ for
$\mathrm{x}\in\left\{\mathrm{error},\mathrm{V},\mathrm{ff},\mathrm{mi}\right\}$ a (dimensionless) transfer function
that describes the bandwidth of various components of the setup as shown in Fig. \ref{fig2}. Only the most limiting
bandwidths are taken into account. The closed loop transfer function can be derived by solving $E(s)$ from Eq.
(\ref{eq8}). Subsequently the error function $e(t)$ can be derived from $E(s)$ by an inverse Laplace transformation.
Although the solution $e(t)$ is analytical, it is not printed here, because it is too lengthy.

The frequency of the laser $\omega_{\mathrm{L}}(t)$ can be derived from the error function by combining the Laplace
transforms of Eqs. (\ref{eq2}), (\ref{eq3}) and (\ref{eq4}) and incorporating the bandwidth transfer functions
$\tau_{\mathrm{x}}(s)$ as discussed previously. This yields for the laser frequency
\begin{equation}
\omega_{\mathrm{L}}(t)=\mathcal{L}^{-1}\left(\frac{E(s)}{\tau_{\mathrm{error}}(s)A}-\tau_{\mathrm{V}}(s)\frac{\delta_1}{s}\right)+\omega_{\mathrm{R}}-\delta_0,
\label{eq9}
\end{equation}
where $\mathcal{L}^{-1}(\cdot)$ denotes an inverse Laplace transformation and the solution of Eq. (\ref{eq8}) for
$E(s)$ should be used for $E(s)$. Also Eq. (\ref{eq9}) yields an analytical but lengthy solution, and is therefore not
printed here. In the next section we will compare the calculated transients with the measured ones.

\section{Comparison with experimental data}
\label{sec4}

\begin{figure}[t]
\centerline{\scalebox{.6}{\includegraphics{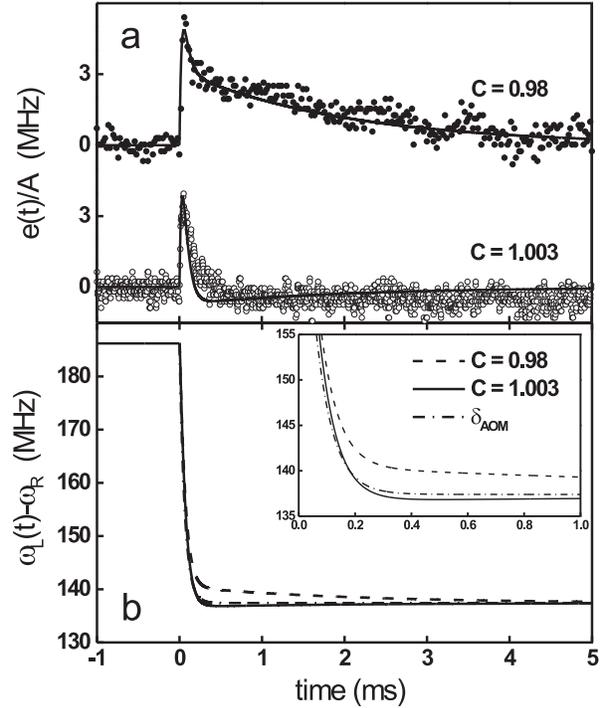}}} \caption{(a) Error signals versus time. At $t=0$ the AOM and
diode laser are switched from MOT detuning to molasses detuning. The top curve shows the error signal and a fit when
the feed-forward compensates 98\% of the applied AOM shift, the bottom curve when the feed-forward is 100.3\% of the
applied AOM shift. The solid line is the prediction of Eq. (\ref{eq8}). (b) Laser frequency versus time for the above
mentioned cases of current feed-forward (solid and dashed), calculated using Eq. (\ref{eq9}), and the frequency shift
of the AOM (dash-dot), which is limited by the bandwidth of the step function $\omega_{\mathrm{V}}$. The inset shows
the development of the laser frequency during the first millisecond in more detail.} \label{fig4}
\end{figure}

In order to lock the laser while the AOM frequency is switching, first the parameter $C$ is coarsely adjusted by
overlapping the spectra as described in section \ref{sec2} (see also Fig. \ref{fig3}b). We then lock the laser to the
desired zero crossing of the error signal by closing the feedback loop. While the laser is locked, the current feedback
parameter $\varphi$ is increased in order to decrease excursions of the error signal. The optimal value of $\varphi$ is
just below the value where the error signal starts to oscillate, in order to be as close as possible to critical
damping. The top curve in Fig. \ref{fig4}(a) shows an error signal when the laser has been locked by this procedure and
the frequency of the laser is changed from MOT to molasses frequencies at $t=0$. The error signal is converted to a
frequency by dividing it by the slope $A$ of the dispersive signal. From this graph it is clear that in steady state
the excursions of the laser frequency are approximately 1 MHz. One recognizes a fast increase of detuning due to the
frequency shift of the AOM, followed by a decrease of frequency shift, because the current feed-forward starts to
compensate the AOM frequency shift. The current feed-forward is slower than the AOM shift due to the bandwidths
$\omega_{\mathrm{ff}}$ and $\omega_{\mathrm{mi}}$. Finally a long tail due to the slow integrational feedback that
removes the last amounts of the error signal is visible. It is clear that the current feed-forward does not completely
cancel the AOM detuning, resulting in a finite error signal that is cancelled by the slow integrational feedback. By
minimizing the amplitude of the error signal while the laser is locked and the AOM detuning is jumping, $C$ can be
optimized to a few permille. A curve with $C$ optimized by using this method is shown in the bottom graph of Fig.
\ref{fig4}(a). In order to determine values for $C$ for both curves, fits to the data using the error signal $e(t)$
derived from Eq. (\ref{eq8}) were performed. The results are shown as solid lines in Fig. \ref{fig4}a. The bandwidths
$\omega_{\mathrm{x}}$ and the frequency $\delta_1$ are kept constant to their measured or specified values. The slope
of the dispersive signal $A$, the integrational feedback parameter $\alpha$ and the current feedback parameter
$\varphi$ are equal for the two curves. Their values are however not accurately known since they include e.g. the
frequency response to the piezo voltage and the laser current. By repeating the fit procedure on both curves while
iteratively varying $A$, $\alpha$ and $\varphi$, these values were optimized. Values for $C$ are determined to be
0.98(1) and 1.003(5) after optimizing using the spectra and error signals respectively. It is clear that with the first
method it is not possible to accurately get $C$ equal to 1, while with the second method this is possible. For both
cases the amplitude of the frequency excursion is smaller than $\Gamma$. We have successfully tested the technique for
frequencies near the extrema of the bandwidth of the AOM. This range thus appears to be the limiting factor. The extent
to which $C$ has to be optimized when the frequency jump is changed depends on the linearity of the response of the VCO
(AOM frequency) and the laser current on the applied voltage. In practice we had to finetune $C$ slightly when the
frequency jump was changed.

Fig. \ref{fig4}(b) shows the frequencies of the laser, calculated using Eq. (\ref{eq9}) for the same parameters as for
the two curves in Fig. \ref{fig4}(a). The frequency shift of only the AOM is also shown. From Fig. \ref{fig4} it is
clear that the error caused by the extra feedback loop is not severe for our application.

The main limiting parameters are the bandwidths $\omega_{\mathrm{mi}}$ and $\omega_{\mathrm{ff}}$ in the current
feed-forward path, which are not present in the electrical path to the AOM. In theory the amplitude of the frequency
excursion can be decreased to 0 by better matching the bandwidths of the two paths, so that the detunings due the AOM
and the current feed-forward path always cancel. It would, of course, be more elegant if the bandwidth
$\omega_{\mathrm{V}}$ were larger, resulting in a shorter step up time. However one should be careful with the
bandwidths of the current feed-forward path and the AOM path, since for constant but unequal bandwidths of these paths
the amplitude of the frequency excursions will increase with $\omega_{\mathrm{V}}$.

\section{Conclusions and outlook}
\label{sec5}

We have demonstrated a new technique for locking a narrow linewidth laser to an arbitrary frequency in the vicinity of
a spectral feature not by frequency shifting the output beam, but by frequency shifting the spectroscopy beam. By
simultaneously switching the AOM frequency and the laser current it is possible to change the frequency of the laser by
more than the locking range, while keeping it locked. Whereas the frequency shifting range in our experiment was
limited by the frequency range of the AOM, it should be possible to make even larger jumps by jumping to a different
lock point. The transient frequency excursion was smaller than 5 MHz, less than the linewidth of the $^{87}$Rb D$_{2}$
transition. The transient time was approximately 0.2 ms. The amplitude of the excursions was limited by the matching of
the bandwidths of the electronics in the feed-forward path and the AOM path. The duration of the transient was limited
by the small bandwidth $\omega_{V}$ of the voltage driving the VCO and the current feed-forward. The demonstrated
technique is not restricted to diode lasers but should be applicable also to other types of laser, e.g. dye or
Ti:Sapphire lasers.

\begin{acknowledgement}
This work is part of the research program of the ``Stichting voor Fundamenteel Onderzoek van de Materie'' (Foundation
for the Fundamental Research on Matter) and was made possible by financial support from the ``Nederlandse Organisatie
voor Wetenschappelijk Onderzoek'' (Netherlands Organization for the Advancement of Research).
\end{acknowledgement}

\end{document}